# Synthesis and properties of charge-ordered thallium halide perovskites, CsTl$^{1+}_{0.5}$Tl$^{3+}_{0.5}$X$_3$ (X = F, Cl)- theoretical precursors for superconductivity?

M. Retuerto[1], T. Emge[1], J. Hadermann[2], P. W. Stephens[3], M. R. Li[1], Z. P. Yin[4], M. Croft[4], A. Ignatov[4], S. J. Zhang[5], Z. Yuan[5], C. Jin[5], J. W. Simonson[3], M. C. Aronson[3,6], A. Pan[7], D. N. Basov[7], G. Kotliar[4], M. Greenblatt[1*]

[1] *Department of Chemistry and Chemical Biology, Rutgers, The State University of New Jersey, 610 Taylor Road, Piscataway, NJ 08854, USA*

[2] *EMAT, Department of Physics, University of Antwerp, Groenenborgerlaan 171, 2020 Antwerp, Belgium*

[3] *Department of Physics & Astronomy, The State University of New York, Stony Brook, NY 11794, USA*

[4] *Department of Physics and Astronomy, Rutgers, The State University of New Jersey, 136 Frelinghuysen Road, Piscataway, NJ 08854, USA*

[5] *Institute of Physics, Chinese Academy of Sciences, Number 8, Zhongguancun South Str 3Beijing 100190, China*

[6] *Brookhaven National Laboratory, Upton, NY 11973, USA*

[7] *Department of Physics, University of California, San Diego, La Jolla, CA 92093-0319, USA*

**Abstract**

Recently CsTlCl$_3$ and CsTlF$_3$ perovskites were theoretically predicted to be potential superconductors if they are optimally doped. The synthesis of these two compounds, together with a complete characterization of the samples are reported. CsTlCl$_3$ is obtained as orange crystals in two different polymorphs: a tetragonal (*I4/m*) and a cubic (*Fm-3m*) phase. CsTlF$_3$ is formed as a light brown powder, also as a double cubic perovskite (*Fm-3m*). In all three CsTlX$_3$ phases Tl$^{1+}$ and Tl$^{3+}$ were located in two different crystallographic positions that accommodate their different bond lengths. In CsTlCl$_3$ some Tl vacancies are found in the Tl$^{1+}$ position. The charge ordering between Tl$^{1+}$ and Tl$^{3+}$ was confirmed by x-ray absorption and Raman spectroscopy. The Raman spectroscopy of CsTlCl$_3$ under high pressure (58 GPa) did not indicate any phase transition to a possible single Tl$^{2+}$ state. However, the highly insulating material becomes less resistive with increasing high pressure, while undergoing a change in the optical properties, from transparent to deeply opaque red, indicative of a decrease of the band gap. The theoretical design and experimental validation of the existence of CsTlF$_3$ and CsTlCl$_3$ cubic perovskites is the necessary first step in confirming the theoretical prediction of superconductivity in these materials.

**Introduction**

The discovery of high-$T_C$ superconductivity in mixed-valent systems such as $La_{2-x}Ba_xCuO_4$ and $Ba_{1-x}K_xBiO_3$,[1-4] have attracted great interest in search for new superconducting materials; in particular in compounds with perovskite-type structure. $BaBiO_3$ becomes superconducting when it is doped as $Ba_{1-x}K_xBiO_3$ ($x \sim 0.4$) and $BaBi_{1-x}Pb_xO_3$ ($x \sim 0.3$).[4,5] In the parent compound $BaBiO_3$, formally $Bi^{4+}$ disproportionates into $Bi^{3+}$ (lone pair $6s^2$) and $Bi^{5+}$ ($6s^0$), which are found at alternating B sites, generating distortions from the ideal perovskite structure. Since the $6s$ orbital has a large radial extension, the Bi-O bond distance is longer for $Bi^{3+}$-O than for $Bi^{5+}$-O.[6] In $Ba_{1-x}K_xBiO_3$ the difference between the two Bi-O bond lengths decreases and superconductivity emerges at the critical $x$, when the structural distortions are completely suppressed.[7] These observations suggest that superconductivity in the $BaBiO_3$ system is mediated by the same mechanism that is responsible for the charge disproportionation. However, superconductivity has not been found in other similar mixed-valent systems. For example, the compounds $CsAuX_3$ (X = Cl, Br, I) with $Au^{1+}$ and $Au^{3+}$ mixed-valent states,[8-10] and exactly one hole per site are semiconductors (at low pressure) with a large band gap. Although, the conductivity of $CsAuX_3$ increases several orders of magnitude at high pressures, superconductivity has not been found thus far.[11-13] In $CsAuX_3$, at ambient pressure, the activation barrier for direct electron pair transfer is very large. Increased pressure forces the geometric differences between different oxidation states to become smaller, decreasing the activation barrier and changing the electronic properties from an insulating - to a metallic state. Although $CsAuX_3$ phases are structurally similar to $BaBiO_3$ and share a similar mixed-valence of the B cations, the absence of superconductivity in the former phases suggests that charge disproportionation is not the sole consideration for superconductivity.

Recently, Kotliar *et al.*[14] explained the high temperature superconductivity in a large family of materials including (Ba,K)BiO$_3$ by a correlation-enhanced strong electron-phonon coupling mechanism. This coupling in $CsAuCl_3$ was found to be weak, accounting for the absence of superconductivity in agreement with experiments. Nevertheless, based on the correlation-enhanced strong electron-phonon coupling mechanism, they predicted that high temperature superconductivity could be found in the doped $CsAuX_3$-like perovskites: $CsTlF_3$ and $CsTlCl_3$.[15] Both of these Tl compounds are expected to be isostructural with $BaBiO_3$, share the same valence electron count and retain essentially the same band structure near the Fermi level. With these similarities, $CsTlF_3$ and $CsTlCl_3$ were predicted to be superconducting with hole doping and pressure to reduce the structural distortions. The critical superconducting temperature ($T_C$) was predicted to be ~30 K for $CsTlF_3$ and ~20 K for the Cl analogue, under an optimal hole doping level (~0.35/f.u.) and moderately high pressures (between ~10 and ~2 GPa, respectively).[15] Recently, similar materials, notably the Rb analogues, have been studied theoretically by Schoop *et al.*,[16] contrasting them with $BaBiO_3$ and predicting the possibility of superconductivity if they were hole doped, and comparing the stability of possible crystal structures upon charge disproportionation.

Here we report the synthesis and properties of $CsTlX_3$ (X= Cl, F) with $Tl^{1+}/Tl^{3+}$ charge ordering. To the best of our knowledge, they are the first inorganic perovskites with charge ordering of Tl cations. A previous synthesis of $CsTlCl_3$ was described in an unpublished dissertation in 1993.[17] We have characterized both materials, determined their crystallographic structures and confirmed $Tl^{1+/3+}$ mixed-valence in both compounds. In addition, we have studied the effect of pressure on the properties. We are currently working on optimally doping these phases with the goal of finding superconducting materials.

**Synthesis Details**

The synthesis of these thallium halides is not trivial due to the toxicity of Tl and the hygroscopic character of the samples. *$CsTlCl_3$* was obtained as fused and intergrown yellow-orange crystals of several millimetres. A precursor of $Cs_2TlCl_5·H_2O$ was prepared by dissolving 3 g of CsCl in 10 ml of HCl (37 %) and 0.3 g of $Tl_2O_3$ in 3 ml of HCl (37 %). When these solutions were combined a precipitate appeared (mainly CsCl); after the addition of 100 ml of water, another precipitate, identified as $Cs_2TlCl_5·H_2O$[18] appeared (Fig. S1a in Section 1 of Supporting Information (SI)). In a second step $Cs_2TlCl_5·H_2O$ + TlCl were mixed and ground inside a glove box, vacuum-sealed in a silica tube and heated to 600 °C/12 h. The powder x-ray diffraction (PXD) pattern of the product (Fig. 1) is characteristic of a distorted tetragonal perovskite. Henceforth, this phase will be designated as $CsTlCl_3$-t. An alternative method of synthesis with TlCl, CsCl and $TlCl_3 \cdot xH_2O$ also yielded $CsTlCl_3$-t by PXD (Figs. S2a and S2b in Section 1 of SI). A different $CsTlCl_3$ phase is obtained when anhydrous $Cs_2TlCl_5$ is used as starting material. ($Cs_2TlCl_5$ is formed by heating $Cs_2TlCl_5 \cdot H_2O$ at 150 $^o$C (Fig. S1b in SI)). When stoichiometric $Cs_2TlCl_5$ + TlCl are ground in the glove box, placed in vacuum-sealed silica tube and heated to 600 $^o$C/12 h. The crystals obtained are lighter yellow-orange (inset of Fig. 2a), but similar in size and shape as the tetragonal phase. This phase has a different structure than the $CsTlCl_3$-t polymorph. It is also perovskite-related but many tetragonal peaks are absent, being explained with a cubic symmetry and designated as $CsTlCl_3$-c (Fig. 2a). Thus we have observed the existence of two different phases in the $CsTlCl_3$ system and a study of both crystal structures is described below. The appearance of different phases appears to be related to the quantity of water in the starting materials; unless they are thoroughly dried, the $CsTlCl_3$-t phase appears. Infrared spectroscopy and thermal gravimetric analysis confirmed the absence of water or hydroxide in either phase (Fig. S3, S4 and S5 in Section 1 of SI).

*CsTlF₃* was also prepared in a glove box to avoid contact of the material with air and moisture. The synthesis is performed in two steps: i) the preparation of TlOF (source of $Tl^{3+}$) by adding $Tl_2O_3$ to a solution of HF (37 %). ii) The reaction $CsF + ½TlF + ½TlOF + xC_2F_4$ ($x$ = 2 to 3). Poly-tetrafluoroethylene ($C_2F_4$) is required as source of F. The starting materials were pressed into a pellet, covered by Pt foil (required to avoid the competing reaction with the silica tube producing $Cs_2SiF_6$) and heated in an evacuated sealed silica tube at 500 $^o$C/12 hs. The obtained material was washed in methanol to eliminate TlCl and CsCl impurities. The Synchrotron powder x ray diffraction (SPXD) pattern is shown in Fig. 3. The experimental details of the techniques used for the characterization of the samples are presented in the Section 1 of the SI.

**Results and Discussion**

We performed chemical analysis by inductively coupled plasma (ICP) mass spectrometry. The composition of $CsTlCl_3$-c, 52.62% Cs: 47.38% Tl, and that of $CsTlCl_3$-t, 53.64 % Cs: 46.36 %

Tl, are very close to the refined (*vide infra*) Tl vacancy concentration: Cs:Tl = 55:45 for CsTlCl$_3$-c (Table 1), and Cs:Tl = 54:46 for CsTlCl$_3$-t (Table 3). For the fluoride analogue it was not possible to obtain a reliable chemical analysis, because of the presence of impurity phases that also contain Cs and/or Tl. However the synchrotron x-ray diffraction data identified all phases present, with the major (~77%) one being the cubic perovskite phase CsTlF$_3$ (see below).

*Crystal Structures*

*CsTlCl$_3$-c:* The crystal structure of CsTlCl$_3$ phases is complex and different techniques were required for their solution. To help solve CsTlCl$_3$-c structure, selected area electron diffraction (SAED) was undertaken. Because the sample decomposed when the electron beam was focussed into a high intensity spot it was not possible to perform imaging (HRTEM, HR-STEM) and convergent-beam electron diffraction (CBED). The unit cell parameters determined by SAED provided sufficient evidence to select the proper space group, *Fm-3m*. All SAED patterns could be indexed with the cell parameters (*a* = 10.84 Å) determined from the PXD data and space group *Fm-3m* (Fig. 4). The refinement of the CsTlCl$_3$-c structure was carried out with single-crystal x-ray diffraction (SCD) data collected at 100 K. The structural refinement was performed with JANA2006.[19] The initial SCD model was based on the known cubic double perovskite, *Fm-3m*, structure with atoms in these positions: Cs$^+$ at A (¼ ¼ ¼), Tl$^{3+}$(Tl1) at B (0 0 0), Tl$^{1+}$(Tl2) at B' (½ 0 0) and Cl$^-$(Cl1) anions at (*x* 0 0) with *x* ~ 0.239(1). The occupancy of Cl$^-$ anion in this site was found to be deficient by about 16% and a second Cl site (Cl2) of approximate occupancy 16% was found 1.9 Å away, at (0 0.165(4) 0.165(4)). Large residuals in the difference-electron-density maps in the vicinity of Tl2 (½ 0 0) and Cs (¼ ¼ ¼) indicated significant disorder around these two positions. We explained the residuals using Gram-Charlier anharmonic displacement parameters up to six orders. The crystallographic details of this refinement are presented in Table S1, the final atomic parameters in Table 1 and the average bond distances in Table 2. The PXD pattern calculated from this refinement (Fig. 2b) is very similar to the experimental PXD data (Fig. 2a). The occupancy of Tl2 in the B' site was found to be less than full, namely 0.68(5) with the highly-correlated intensity weighting scheme based upon p = 0.10 for [σ(I) + p*I$^4$] in JANA2006. Thus, the B' site appears to contain some vacancies and requires the presence of both $Tl^{1+}$ and $Tl^{3+}$ in this position for charge balance, but is mainly occupied by $Tl^{1+}$. This result would explain the shorter Tl2-Cl1 distances found (2.83(1) Å) compared with the expected bond distance of 3.31 Å from the $Tl^{1+}$-Cl Shannon radii sum,[20] while Tl1-Cl distances match very well with six-coordinate $Tl^{3+}$ cations, 2.70 Å. Occurrences of vacancies in other perovskite-like chlorides ($Cs_2MCl_6$ with $K_2PtCl_6$ structural type, *Fm-3m*) have been reported, and take place when M is a tetravalent cation, such as Sn, Ta, Re, W, etc.[21-24] 12-coordinated $Cs^{1+}$ is primarily bonded to Cl1 (0.84 refined occupancy for Cl1) at distances of 3.828(1) Å, which agrees well with the Shannon radii sum (3.69 Å). There are much fewer (~0.16 of the time) Cs to Cl2 bonds with smaller distances (2.97(2) Å), which are, probably due to some $Tl^{1+}$ at this position or vacancies, consistent with the refined stoichiometry. We also investigated the possible existence of oxygen over either Cl$^-$ site, however, $O^{2-}$ did not refine as well as Cl$^-$ and gave unrealistic metal-oxide bond lengths. Further examination of the two Cl$^-$ sites reveals that only Cl1 atoms are close enough to Tl2 to form bonds with it; but both Cl1 and Cl2 form bonds (but not at the same time) with Cs and Tl1 (Table 2). The appearance of an interstitial anion (as Cl2 in our case) has been observed in some oxygen-containing double perovskites with large B-site cations (as $Tl^{1+}$ in our case), such as $Sr_2(Sr_{1-x}M_{1+x})O_6$ (B = Nb, Ta) and $Sr_2MSbO_6$ (M = Ca, Sr, Ba).[25-27] The large B-site cation causes non-harmonic atomic displacements that drives some oxygen to interstitial site and leaves oxygen vacancies in the original positions.

Another model with split positions for the atom (low symmetry disorder of the cations about their high symmetry sites) in program SHELXL[28] was also used to explain the non-spherical electron density and better quantify the apparently anharmonic displacements of Cs and Tl atoms about sites A and B', respectively. Partially-weighted Cs atoms were placed at the A site and also at 2 unique sites nearby, yielding, after application of the site symmetry, a regular polygon with 10 vertices of partially-weighted Cs atoms around the central Cs atom at (¼ ¼ ¼) with average distances from (¼ ¼ ¼) of approximately 0.5 Å. A similar procedure for Tl2 in the B' site yielded a regular polygon with 18 vertices of Tl atoms spaced about 0.9 Å from the central Tl at (½ 0 0). In this model, the occupancy sum of all partially-weighted Tl2 in the B' site was found to be less than full. The crystallographic details of the split-atom refinement are shown in Table S2. The atomic and anisotropic displacement parameters are presented in Table 3, and the bond distances in Table 4. The PXD pattern calculated from this split atom refinement is illustrated in Fig. 2c, together with a schematic view of the proposed structure. This type of refinement has been previously reported in other, similarly disordered systems, such as $Ni_{1+\delta}Sn$ or $LaNi_5$.[29,30] The obtained distances with this model are similar to that of the Gram-Charlier anharmonic displacement model. The Cs1-Cl1, Cs2-Cl1 and Cs3-Cl1 distances in the range 3.41-to-3.87 Å are comparable to the Shannon sum of 3.69 Å for 12-coordinate Cs, while for 0.16 of the time the shorter Cs-Cl2 indicated for Cs1, Cs2 and Cs3 with Cs-Cl in the range 2.56-to-3.46 Å and a coordination number of 6; both observations are consistent with the presence of $Tl^{1+}$ or vacancies at these Cs positions. The Tl1-Cl1 and Tl1-Cl2 distances are all at 2.57 Å for the 6-coordinate Tl, and agree well with the expected (Shannon) $Tl^{3+}$-Cl distance of 2.69 Å. The triply-disordered Tl2 distances appear to be approximately 6- or 4- or 2-coordinate for Tl2A (six Tl2A-Cl1 are 2.84 Å), Tl2B (four Tl2B-Cl1 are 2.98 Å, and the one 1.96 Å distance is ignored as a vacancy of Tl2B) or Tl2C (two Tl2C-Cl1 are 2.98 Å, and the two 2.30 Å distances are ignored as vacancies of Tl2C), respectively. As in the anharmonic displacement model, these distances are only to Cl1 and not Cl2 and are always shorter than the expected (Shannon) 3.31 Å distance for $Tl^{1+}$-Cl, and most likely can be explained by the presence of some $Tl^{3+}$ or vacancies. For comparison, the main $Cs^{1+}$-Cl (~ 3.82 Å) and $Tl^{3+}$-Cl (~ 2.58 Å) distances obtained in both models are comparable with those in similar compounds including $Cs_2TlCl_5$ (Cs-Cl: 3.6816 Å and $Tl^{3+}$-Cl: 2.5963 Å)[31] and $Tl_2Cl_3$ ($Tl^{3+}$-Cl~ 2.57 Å)[32]. $Tl^{1+}$-Cl distances are difficult to compare with other compounds since we likely have some vacancies and $Tl^{3+}$ in those positions and also $Tl^{1+}$ in relatively low (i.e., ≤ 6) coordination is rare.

*CsTlCl₃-t:* Several tilt series of SAED data were taken from different crystallites to determine the unit cell. The compound showed clear superstructure reflections, as was evidenced from the two most prominent zones shown in Fig. 5. To index all the reflections, a supercell was needed, with cell parameters approximately $a \approx b \approx 17.0$ Å, $c \approx 11.0$ Å. The relationship between this supercell and the cell of a simple tetragonal double perovskite with subcell a ≈ b ≈ 7.7 Å, c ≈ 11.0 Å, is given by the transformation matrix $P=\begin{bmatrix} 1 & -2 & 0 \\ 2 & 1 & 0 \\ 0 & 0 & 1 \end{bmatrix}$, which yields a 5-fold increase in the unit cell volume and a resultant tetragonal superlattice of $5^{1/2}$**a** $5^{1/2}$**b c**. From the analysis of the reflection conditions and intensity distributions in the ED and synchrotron powder x-ray diffraction (SPXD) patterns, combined with prediction of the probable space group by SPUDS,[33] space group *I4/m* was derived. An initial model for the structure was obtained from precession electron diffraction data (PED). The details of the PED experiment and the structure solution from the PED data can be found in Section 3 of the SI. The model was ultimately refined with SPXD. In this structure there are 2 positions for Cs, 4 for Tl (being Tl1 and Tl3 mainly $Tl^{3+}$ and Tl2 and Tl4, $Tl^{1+}$) and 7 for Cl. Sites Tl2 and Tl4 have occupancy significantly less than unity. A schematic view of the structure together with the refinement of the structure with SPXD are illustrated in Fig. 1. The atomic positions are shown in Table 5 and the bond distances in Table 6. The structure is basically formed by corner-sharing $TlCl_6$ octahedra ($O_h$) with Cs in the voids. Some of the $TlCl_6$ $O_h$ are highly distorted producing a strong tilting of other $O_h$. The reason for these distortions could be related to the lone pair effect of the $Tl^{1+}$ cation together with its large size. For example, $Tl4Cl_6$ are highly distorted from ideal $O_h$, producing a tilting of $Tl1Cl_6$ of ~45° relative to the orientation of $Tl2Cl_6$ and $Tl3Cl_6$ (inset of Fig. 1). The Tl1-Cl and Tl3-Cl bond distances are consistent with what is expected for $Tl^{3+}$-Cl. The Tl2-Cl and Tl4-Cl distances are consistent with, but slightly shorter than, what is expected for $Tl^{1+}$-Cl; this finding is likely due to the presence of some vacancies and/or some $Tl^{3+}$ over the Tl2 and Tl4 positions. For this set of Tl sites, that are either $Tl^{3+}$ (Tl1 and Tl3) or mostly $Tl^{1+}$ (Tl2 and Tl4), we confirm that there is charge ordering in the structure. It is also worth noting that the dramatic superlattice of $5^{1/2}$**a** $5^{1/2}$**b c** determined for this phase is reported for the first time in a perovskite-related material.

*CsTlF₃:* The crystallographic structure was refined from SPXD data with *Fm-3m* space group ($R_{wp}$= 9.8%) and the refined cell parameters was *a* = 9.5449(1) Å, the position of F was (0.277(1), 0, 0), and the observed $Tl^{3+}$-F and $Tl^{1+}$-F distances were 2.12(1) and 2.65(1) Å, respectively, consistent with $Tl^{3+}$-F and $Tl^{1+}$-F distances in other phases, including $Tl^{3+}$ in $Cs_3TlF_6$ (2.02 Å)[34] and $Tl^{1+}$ in TlF (2.4-2.7 Å).[35] The Tl-F distances are also comparable with those expected from the ionic radii sums of 2.21 Å for $Tl^{3+}$- F and 2.83 Å for $Tl^{1+}$-F.[20] Fig. 3 shows the refinement of the structure in the *Fm-3m* cubic model with the presence of the impurities: $Cs_3TlF_6$ (*I4/mmm*) 14 wt %, $Tl_2O_3$ (*Ia-3*) 6 %, $CsO_2$ (*I4/mmm*) 2 % and TlF (*Pm2a*) 1

%. The schematic view of the structure of $CsTlF_3$ is shown in the inset of Fig. 3. The discovery of cubic structures in $CsTlX_3$ is a good starting point in search of superconductivity since, in theory, superconductivity in these phases could only be achieved in cubic perovskites.[15]

*X-ray Absorption Spectroscopy (XAS)*

We performed XAS to verify the mixed valence character of thallium, which commonly occurs in $Tl^{1+}$ and $Tl^{3+}$ valence states; $Tl^{3+}$ involves two 6s-orbital holes. The $Tl-L_3$ edge XAS signature of $Tl^{1+}$ to $Tl^{3+}$ change is a chemical shift of the main edge to higher energy and the appearance of a shoulder pre-edge feature due to transitions into the empty 6s hole states.[36] These same signatures are well known at the $Bi-L_3$ edge to evidence the $Bi^{3+}$ to $Bi^{5+}$ change.[37-39] In Fig. 6a the $Tl-L_3$ edge spectra of $CsTlCl_3$-t and $CsTlCl_3$-c are compared to those of $Tl^{1+}$ and $Tl^{3+}$ standards. Assigning the nominal chemical shift as the energy where the normalized absorption coefficient value first rises to the $\mu = 0.5$ value; the chemical shifts of $CsTlCl_3$ phases are clearly intermediate between those of $Tl^{1+}$ and $Tl^{3+}$ standards supporting their averaged valence. The disparity between the energies of the $\mu = 0.5$, chemical shift points of the $Tl^{3+}$ standards, $Cs_2TlCl_5$ and $Tl_2O_3$ underscore the qualitative nature of the inference of the intermediate valence of both $CsTlCl_3$ phases. In Fig. 6b the pre-edge region of the spectra are plotted on an expanded scale with the $CsTlCl_3$-t spectrum being left out since it is essentially identical to that of $CsTlCl_3$-c. The first point to note is that the $Tl^{1+}Cl$ spectrum exhibits a monotonic concave upward curvature over the entire pre-edge region, consistent with the absence of any 6s hole states ($6s^2$). In contrast the $Tl^{3+}$ standards manifest pronounced pre-edge shoulder features consistent with the presence of two 6s-orbital holes ($6s^0$). $CsTlCl_3$-c shows a clear, albeit not dramatic, shoulder feature supporting the presence of 6s hole states and a $Tl^{1+}/Tl^{3+}$ valence state. To emphasize the presence of the shoulder, a sp line fit to the data above and below this feature has been subtracted, to obtain an estimate of just the shoulder feature portion ($\Delta\mu$), shown in the bottom of Fig. 6b. The $Tl-L_3$ edge of $CsTlF_3$ and the $Cl-K$ edge of $CsTlCl_3$ phases are presented in Section 4 of the SI.

*Raman Spectroscopy*

We have studied the effect of pressure on the Raman spectra of CsTlCl$_3$ phases (Fig. 7) to investigate a possible transition to a single valence state (Tl$^{1+}$/Tl$^{3+}$-to-Tl$^{2+}$). Further, we carried out frozen phonon calculations with hybrid Density Functional Theory (DFT) for the Tl-Cl stretching mode of cubic CsTlCl$_3$ and the calculated phonon frequency was found to be 277 cm$^{-1}$, in excellent agreement with that observed at 270 cm$^{-1}$ (Fig. 7). With increasing pressure, the 270 cm$^{-1}$ Raman Tl-Cl mode moves to higher frequency (as expected due to the compression of the bonds) but does not completely disappear, as is the case of CsAuX$_3$, where Au$^{1+/3+}$ charge ordering is suppressed at a critical pressure.[40] This result does not rule out the possibility of the compound becoming superconducting with optimal doping, since a similar effect was observed in undoped BaBiO$_3$, where a single valence state by merely applying high pressure, without doping, was not possible to achieve.[41] The pressure dependent Raman spectra of CsTlCl$_3$-t indicate similar behaviour.

*Resistivity Measurements under pressure*

We measured the resistance of CsTlCl$_3$-c at high pressures to study a possible transition to a metallic state. The resistance is too high to be measurable below 50 GPa. Fig. 8 shows the resistance *vs.* temperature at 53, 58 and 60 GPa; CsTlCl$_3$-c becomes more conductive and with a change from transparent to opaque with increasing pressure that indicate the narrowing of the band gap. The activation energy $E_g$, which is proportional to the energy gap, and can be calculated from the relation "lnR ~ $E_g/2k_BT$", was found to be 0.245, 0.233, and 0.212 eV for 53, 58 and 60 GPa, respectively. These values are one order of magnitude smaller than the optical band gap predicted theoretically (2.1 eV; Fig.9) and measured experimentally (2.5 eV; Inset of Fig. 9) for CsTlCl$_3$-c without pressure. These results indicate that the band gap is decreasing with pressure, in good agreement with theory. The magnetic behavior of the materials reflects essentially insulating/diamagnetic properties as discussed in the Section 5 of the SI.

*First-principles calculations*

One of the important characteristics of the compounds is that the semi-local local density approximation (LDA) and generalized gradient approximation (GGA) within the density functional theory (DFT) framework substantially underestimate the structural distortions and band gaps; and incorporating non-local correlations (long-range exchange) with HSE-like (*screened*) hybrid functional brings both quantities to much better agreement with experiments.[14,15] To test this, we carried out first-principles DFT calculations with the VASP code[42] for $CsTlCl_3$-c and compare the results with available experiments. In addition to a generalized gradient approximation (PBE version)[43] exchange-correlation functional, a *screened* hybrid functional HSE06[44] has been adopted to account for the non-local correlation/long-range exchange in this compound. Using the experimental lattice constant $a$ = 10.8226(14) Å for $CsTlCl_3$-c, we relax the structure within the *Fm-3m* space group. The relaxed Cl position ($x$ 0 0) is $x$ = 0.2392 from DFT-PBE and $x$ = 0.2359 from DFT-HSE06. Comparing with the experimental value of $x$ = 0.2376(10), DFT-PBE underestimates the Cl breathing distortion while DFT-HSE06 slightly overestimates it, similar to the case of $BaBiO_3$. Likewise, the DFT-HSE06 optimized value $x$ = 0.226 of the F position ($x$ 0 0) in $CsTlF_3$ at ambient pressure agrees better with the experimental value $x$ = 0.223(1) (equivalent to 0.277(1) in a different setting) than the DTF-PBE optimized value $x$ = 0.230 (see Section 6 of the SI). Adopting the experimental crystal structure, we compute the total density of states (DOS) and optical conductivity of $CsTlCl_3$-c (Figs. 9a and b, respectively): $CsTlCl_3$-c is an insulator with a(n) (indirect) band gap of about 1.3 eV (0.6 eV) and an optical gap of about 2.1 eV (1.4 eV), as computed by DFT-HSE06 (DFT-PBE). The calculated band gap by DFT-GGA is much smaller than the DFT-HSE06 value, illustrating the delocalization error of the semi-local nature of the LDA/GGA functional. This delocalization error also has an impact on the positions of the Cl $p$ states, and consequently the peak positions in the optical conductivity. As shown in Fig. 9a, the valence bands near the Fermi level are dominated by Cl $p$ states, peaking at about -3.0 (-2.4) eV in DFT-HSE06 (DFT-GGA). Correspondingly, the first peak position in the optical conductivity is at about 2.1 (1.4) eV (Fig. 9b). The DFT-HSE06 optical gap agrees reasonably well with the experimental optical gap of about 2.5 eV estimated from the optical experiments in transmission mode (Inset Fig. 9b), and is much larger than the result of the corresponding DFT-GGA calculation using the same crystal structure. The above results confirm that CsTlCl$_3$-c behaves the same way as BaBiO$_3$, and could be another member of the family of the "other high temperature superconductors".[14]

**Conclusions**

In conclusion, we have synthesized Tl$^{1+}$/Tl$^{3+}$ mixed- valent CsTlX$_3$ (X = F, Cl) materials, which were theoretically predicted to become superconducting with appropriate doping and under pressure. This is the first published report of an inorganic perovskite with Tl$^{1+}$/Tl$^{3+}$ charge ordering stabilized. CsTlCl$_3$ intergrown transparent yellow-orange crystals were obtained as two phases: a distorted tetragonal (*I4/m*) and a cubic (*Fm-3m*) phase with interstitial Cl$^-$ balancing Cl$^-$ site vacancies and some Tl$^{1+}$ vacancies. CsTlF$_3$ is obtained as a brown polycrystalline material with cubic symmetry (*Fm-3m*). XAS and Raman spectroscopy unambiguously confirm the presence of Tl$^{1+}$-Tl$^{3+}$. Raman spectra demonstrate that the charge ordering is present even at high pressures. The resistance of the cubic CsTlCl$_3$ is too high to measure under ambient conditions, but it decreases, when high pressures are applied. First-principles calculations of the crystal structure, density of states and optical conductivity, together with the corresponding experimentally determined properties of cubic CsTlCl$_3$ indicate that it is similar to BaBiO$_3$, and a potential superconductor, when it is optimally doped.

**Supporting Information Available**

Details of PXD of the precursors and different obtained materials, TGA and DSC analysis, IR spectroscopy. Experimental details. Crystallographic details of the two models used to explain the cubic CsTlCl$_3$-c phase. Details of the structure solution from PED data, including initial model. XAS Tl-K edge of CsTlF$_3$ and Cl-K edge of CsTlCl$_3$ phases. Magnetic Measurements. The density of states (DOS) and optical conductivity of CsTlF$_3$.

**Acknowledgement**

This work was supported by NSF-DMR-0966829, DOD-VV911NF-12-1-0172 and Rutgers University (BOG) grants. Z. P. Y. and G. K. were supported by the AFOSR-MURI program towards better and higher temperature superconductors. Use of the National Synchrotron Light Source, Brookhaven National Laboratory, was supported by the U.S. Department of Energy, Office of Science, Office of Basic Energy Sciences, under Contract No. DE-AC02-98CH10886. We want to thank Dr. Hongbing Sun and James Elliott for their help with the ICP and IR measurements.

**Table 1.** Atomic positions, anharmonic displacement parameters and site occupancies for CsTlCl$_3$-c refined in the F*m-3m* space group (No. 225), cell parameter a = 10.8226(14) Å.

| Atom | Cs | Tl1 | Tl2 | Cl1 | Cl2 |
|---|---|---|---|---|---|
| x | ¼ | 0 | ½ | 0.2376(10) | 0 |
| y | ¼ | 0 | 0 | 0 | 0.165(4) |
| z | ¼ | 0 | 0 | 0 | 0.165(4) |
| Occ(Fraction) | 1.000 | 1.000 | 0.68(5) | 0.84(4) | 0.16(4) |
| $U_{11}$ | 0.092(11) | 0.0245(11) | 0.151(7) | 0.028(5) | 0.03(2) |
| $U_{22}$ | 0.092(11) |  | 0.151(7) | 0.049(4) | 0.03(2) |
| $U_{33}$ | 0.092(11) |  | 0.151(7) | 0.049(4) | 0.03(2) |
| $U_{12}$ | 0 |  | 0 | 0 | 0 |
| $U_{13}$ | 0 |  | 0 | 0 | 0 |
| $U_{23}$ | 0 |  | 0 | 0 | 0.01(2) |
| $C_{123}$ | +0.007(7) |  |  |  |  |
| $D_{1111}$ | -0.007(5) |  | -0.03(3) |  |  |
| $D_{1122}$ | 0.000(3) |  | -0.071(10) |  |  |
| $E_{11123}$ | -0.0004(7) |  |  |  |  |
| $F_{111111}$ | -0.004(2) |  | -0.08(2) |  |  |
| $F_{111122}$ | -0.0002(4) |  | -0.013(2) |  |  |
| $F_{112233}$ | +0.0004(4) |  | +0.011(4) |  |  |

**Table 2.** Bond lengths [Å] for CsTlCl$_3$-c from refinement in JANA2006 using anharmonic displacement parameters.

| | |
|---|---|
| Cs-Cl1 | 12x 3.828(1) |
| Cs-Cl2 | 12x 2.97(2) |
| Tl1-Cl1 | 6x 2.58(1) |
| Tl1-Cl2 | 6x 2.59(5) |
| Tl1-Cl2 | 6x 2.59(5) |
| Tl2-Cl1 | 6x 2.83(1) |

The values of the distances for the cation positions, Cs and Tl2 are average values with maximum displacements from their site centres of 0.5 and 0.9 Å, respectively (see text).

**Table 3.** Atomic coordinates, occupancies and displacement parameters ($Å^2$) for $CsTlCl_3$-c, with the split atom model. The isotropic displacement parameters U(eq) are defined as one third of the trace of the orthogonalized $U^{ij}$ tensor. The anisotropic displacement factor exponent takes the form: $-2\pi^2[ h^2 a^{*2} U^{11} + ... + 2 h k a^* b^* U^{12} ]$.

| Atom | Cs1 | Cs2 | Cs3 | Tl1 | Tl2A | Tl2B | Tl2C | Cl1 | Cl2 |
|---|---|---|---|---|---|---|---|---|---|
| $x$ | ¼ | 0.2791(10) | ¼ | 0 | ½ | 0.4180(3) | 0.4414(12) | 0.2373(2) | 0 |
| $y$ | ¼ | 0.2209(10) | ¼ | 0 | 0 | 0 | 0 | 0 | 0.1682(8) |
| $z$ | ¼ | 0.2209(10) | 0.2059(12) | 0 | 0 | 0 | -0.586(12) | 0 | 0.1682(8) |
| Occ(Fraction) | 0.30(1) | 0.30(2) | 0.40(3) | 1.000 | 0.026(2) | 0.62(1) | 0.13(1) | 0.82(3) | 0.18(3) |
| U(eq) | 0.038(1) | 0.038(1) | 0.038(1) | 0.024(1) | 0.024(1) | 0.024(1) | 0.024(1) | 0.042(1) | 0.034(3) |
| $U_{11}$ | | | | | 0.038(2) | 0.038(2) | 0.038(2) | 0.028(1) | 0.033(5) |
| $U_{22}$ | | | | | 0.017(1) | 0.017(1) | 0.017(1) | 0.048(1) | 0.034(4) |
| $U_{33}$ | | | | | 0.017(1) | 0.017(1) | 0.017(1) | 0.048(1) | 0.034(4) |
| $U_{12}$ | | | | | 0.02(8) | 0.02(8) | 0.01(10) | 0 | 0 |
| $U_{13}$ | | | | | -0.01(6) | -0.01(6) | -0.03(5) | 0 | 0 |
| $U_{23}$ | | | | | 0.013(3) | 0.013(3) | 0.03(10) | 0 | -0.013(5) |

**Table 4.** Bond lengths [Å] for $CsTlCl_3$-c, using the split atom model.

| | |
|---|---|
| Cs1-Cl1 | 12x 3.8288(5) |
| Cs1-Cl2 | 1x 2.981(5) |
| Cs1-Cl2 | 6x 2.981(5) |
| Cs2-Cl1 | 6x 3.856(2), 3x 3.412(14) |
| Cs2-Cl2 | 3x 2.735(9), 3x 3.126(7) |
| Cs2-Cl2 | 3x 2.735(9), 3x 3.464(18) |
| Cs3-Cl1 | 4x 3.508(9), 2x 3.8415(13), 2x 3.875(2) |
| Cs3-Cl2 | 2x 2.555(13), 4x 3.156(9) |
| Cs3-Cl2 | 2x 3.421(13), 4x 2.876(4) |
| Tl1-Cl1 | 6x 2.569(2) |
| Tl1-Cl2 | 6x 2.575(13) |
| Tl1-Cl2 | 6x 2.575(13) |
| Tl2A-Cl1 | 6x 2.843(2) |
| Tl2B-Cl1 | 4x 2.978(2), 1x 1.955(4) |
| Tl2C-Cl1 | 2x 2.297(9), 2x 2.981(6) |

**Table 5.** Unit-cell, positional, displacement parameters and site occupancies for CsTlCl$_3$-t refined from SXPD in the I4/m space group (No. 87), and cell parameters a = 17.2489(2) Å and c = 11.1004(2) Å. Agreement factors: R$_p$= 5.15%, R$_{exp}$= 3.51%, R$_{wp}$= 6.61%, R$_{Bragg}$= 3.1 %, and $\chi^2$ = 3.54.

| Atom | Wyckoff | x | y | z | Occ(Fraction) | Uiso |
|---|---|---|---|---|---|---|
| Cs1 | 4d | 0 | 0.5 | 0.25 | 1 | 0.07(1) |
| Cs2 | 16i | 0.2138(3) | 0.1117(2) | 0.2750(4) | 1 | 0.07(1) |
| Tl1 | 2a | 0 | 0 | 0 | 1 | 0.032(1) |
| Tl2 | 2b | 0 | 0 | 0.5 | 0.61(1) | 0.032(1) |
| Tl3 | 8h | 0.0942(3) | 0.3023(2) | 0.5 | 1 | 0.0315(6) |
| Tl4 | 8h | 0.0840(3) | 0.2568(2) | 0 | 0.82(1) | 0.031(6) |
| Cl1 | 4e | 0 | 0 | 0.231(3) | 1 | 0.071(2) |
| Cl2 | 16i | 0.087(1) | 0.295(1) | 0.265(1) | 1.000 | 0.071(2) |
| Cl3 | 8h | 0.122(1) | 0.073(2) | 0 | 1 | 0.071(2) |
| Cl4 | 8h | 0.135(2) | 0.453(3) | 0.5 | 1 | 0.071(2) |
| Cl5 | 8h | 0.253 (2) | 0.259(2) | 0.5 | 1 | 0.071(2) |
| Cl6 | 8h | −0.059(1) | 0.339(2) | 0.5 | 1 | 0.071(2) |
| Cl7 | 8h | 0.048(2) | 0.172(2) | 0.5 | 1 | 0.071(2) |

**Table 6.** Bond lengths [Å] for CsTlCl$_3$-t.

| | |
|---|---|
| Tl1 – Cl3 | 4x 2.45(3) |
| Tl1 – Cl1 | 2x 2.56(3) |
| Tl2 – Cl1 | 2x 2.99(3) |
| Tl2 – Cl7 | 4x 3.07(3) |
| Tl3 – Cl7 | 2.39(3) |
| Tl3 – Cl2 | 2x 2.62(1) |
| Tl3 – Cl4 | 2.64(3) |
| Tl3 – Cl6 | 2.71(3) |
| Tl3 – Cl5 | 2.84(3) |
| Tl4 – Cl5 | 2.82(3) |
| Tl4 – Cl4 | 2.93(3) |
| Tl4 – Cl2 | 2x 3.01(1) |
| Tl4 – Cl3 | 3.26(3) |
| Tl4 – Cl6 | 3.45(3) |
| Minimum Tl – Cs | 4.53(1) |
| Minimum Cl – Cs | 3.44(1) |

**Bibliographic References**

**Figure captions**

Fig. 1: Refinement of the structure of CsTlCl$_3$-t using SPXD. Lattice parameters $a$ = 17.2489(2) Å and $c$ = 11.1004(2) Å. Space group *I4/m*. Inset: schematic view of the structure (Tl$^{1+}$ inside the green octahedra, Tl$^{3+}$ inside orange ones and Cs$^{1+}$ are the grey spheres).

Fig. 2: Comparison of (a) experimental PXD of CsTlCl$_3$-c. Inset: Crystals of CsTlCl$_3$-c. (b) Simulation from SCD using whole-site anharmonic ADP model. (c) Simulation from SCD using the split atom model. The inset shows a schematic view of the structure with this model. The green balls represent the disordered Tl$^{1+}$ atoms and the orange ones Tl$^{3+}$ in regular O$_h$ site.

Fig. 3: Rietveld refinement of CsTlF$_3$ in *Fm-3m* space group with a = 9.5449(1) Å, using SPXD. Black dots are raw data, the red line is the computed model of CsTlF$_3$, but all phases were included in calculation of the difference curve. Inset: Cubic double perovskite structure of CsTlF$_3$.

Fig. 4: Representative selected area electron diffraction patterns of CsTlCl$_3$-c, indexed in the F-centered cell described in the text.

Fig. 5: Three most prominent zones from the series taken from CsTlCl$_3$-t. Patterns are indexed in the supercell mentioned in the text.

Fig. 6: (a) Tl-L$_3$ edges of CsTlCl$_3$–t, CsTlCl$_3$-c, and Tl$_2$O$_3$, Cs$_2$TlCl$_5$ and TlCl standards. (b) An expanded view of the Tl-L$_3$ pre-edges. The 6s-hole related feature of CsTlCl$_3$-c (extracted by subtraction of a sp line fit background) is also shown on an enlarged vertical scale in the bottom of the figure.

Fig. 7: Pressure dependence of the Raman spectra of CsTlCl$_3$-c at room temperature.

Fig. 8: Resistance *vs.* Temperature of CsTlCl$_3$-c measured under 53, 58 and 60 GPa applied pressures (measuring current I 0.1uA); inset shows resistance *vs.* pressure at several fixed temperatures.

Fig 9: (a) Total density of states (DOS). (b) Calculated optical conductivity of CsTlCl$_3$-c. Inset: Measured optical conductivity of CsTlCl$_3$-c and CsTlCl$_3$-t.

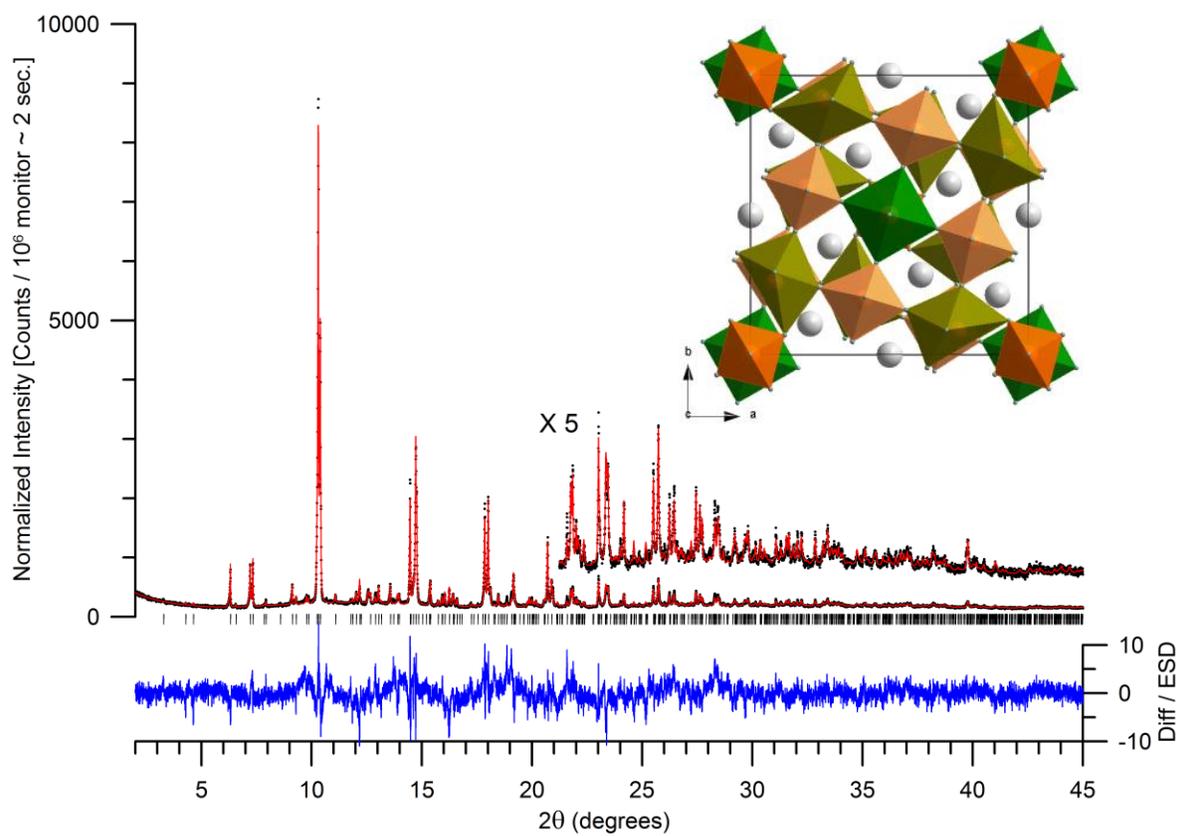

Fig. 1

Retuerto *et al.*

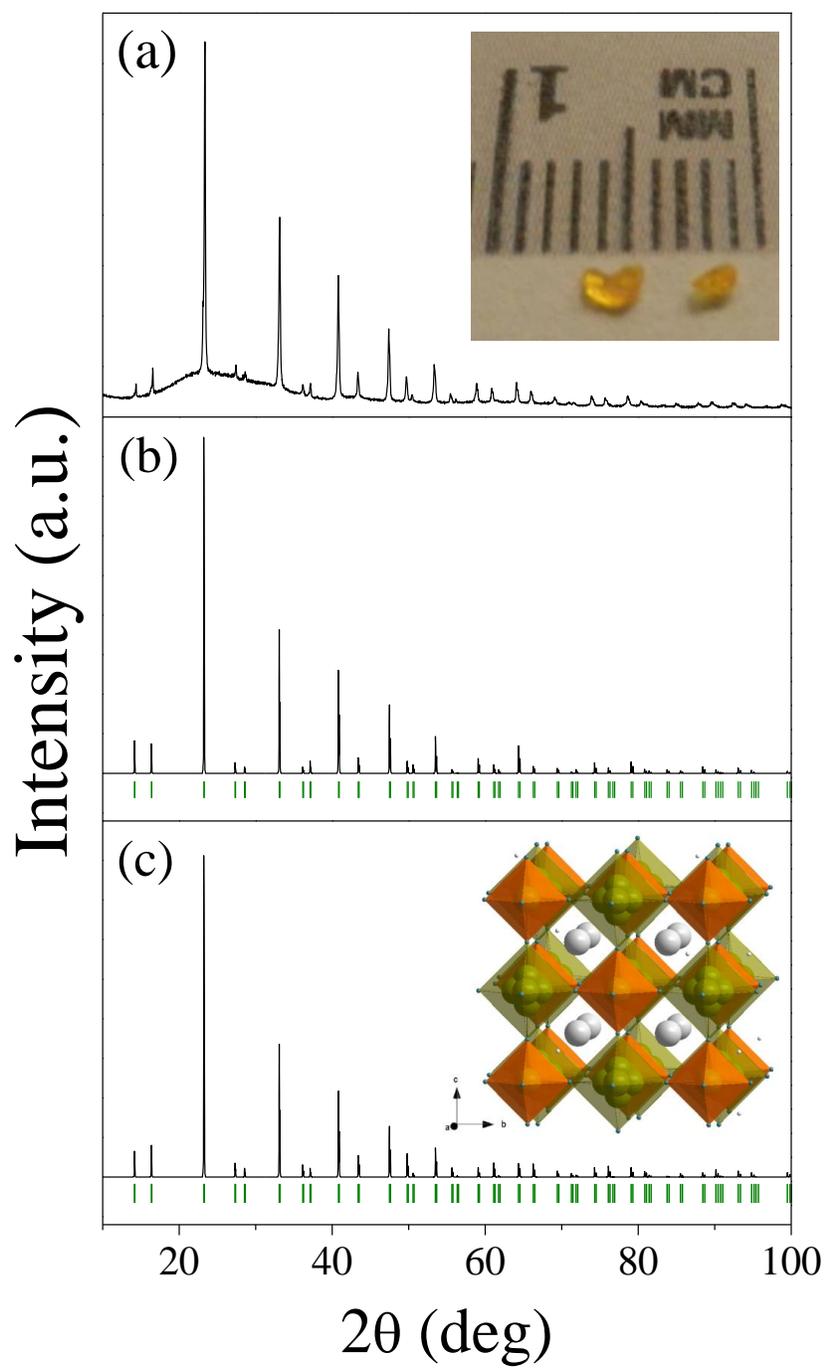

Fig. 2

Retuerto *et al*.

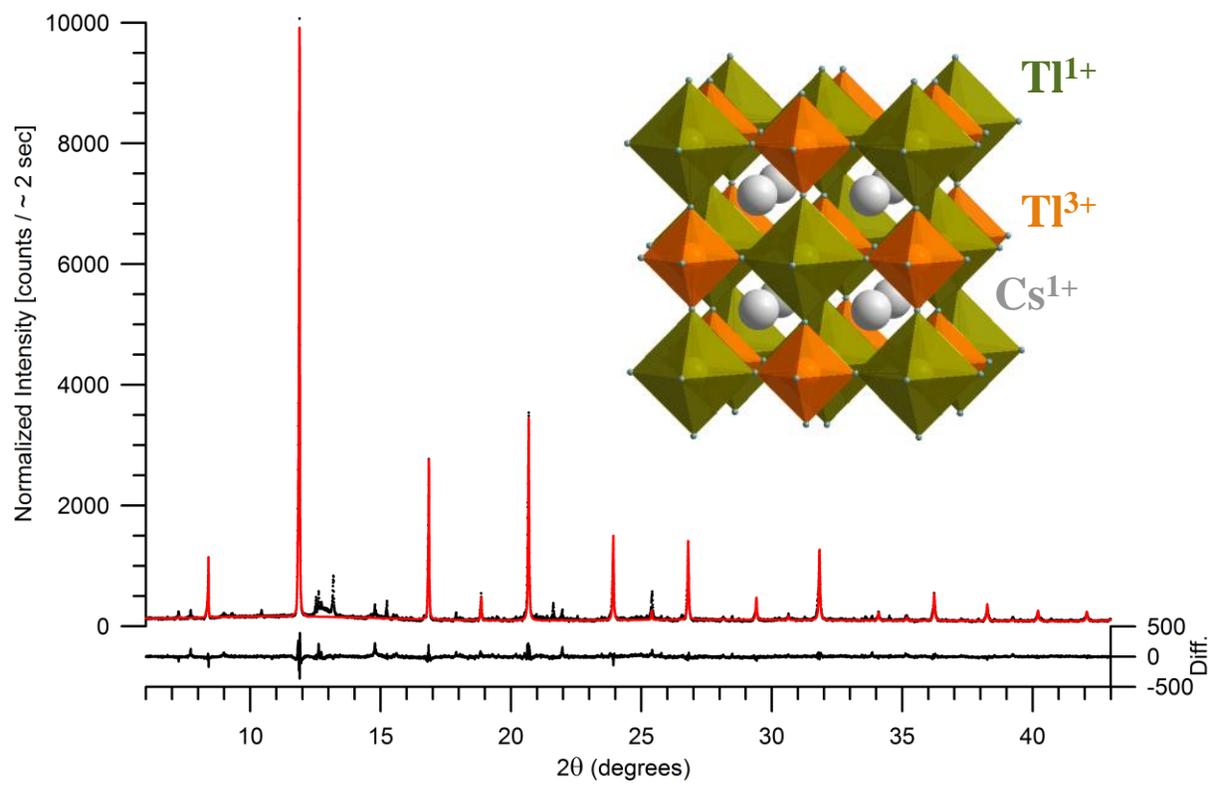

Fig. 3

Retuerto *et al.*

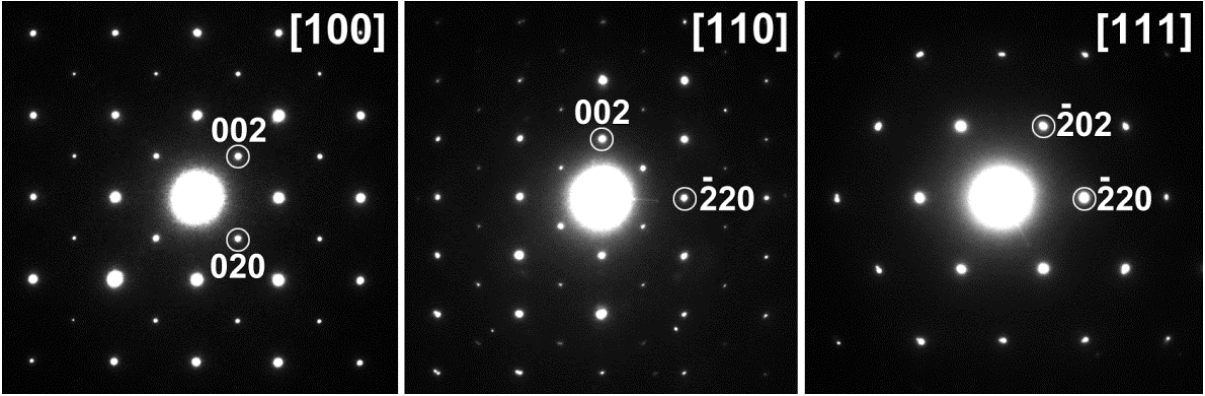

Fig. 4

Retuerto *et al.*

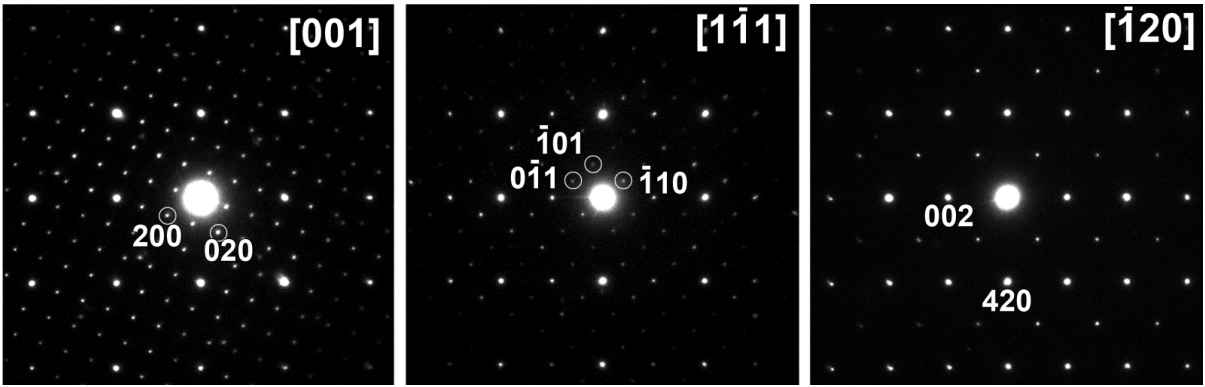

Fig. 5

Retuerto *et al.*

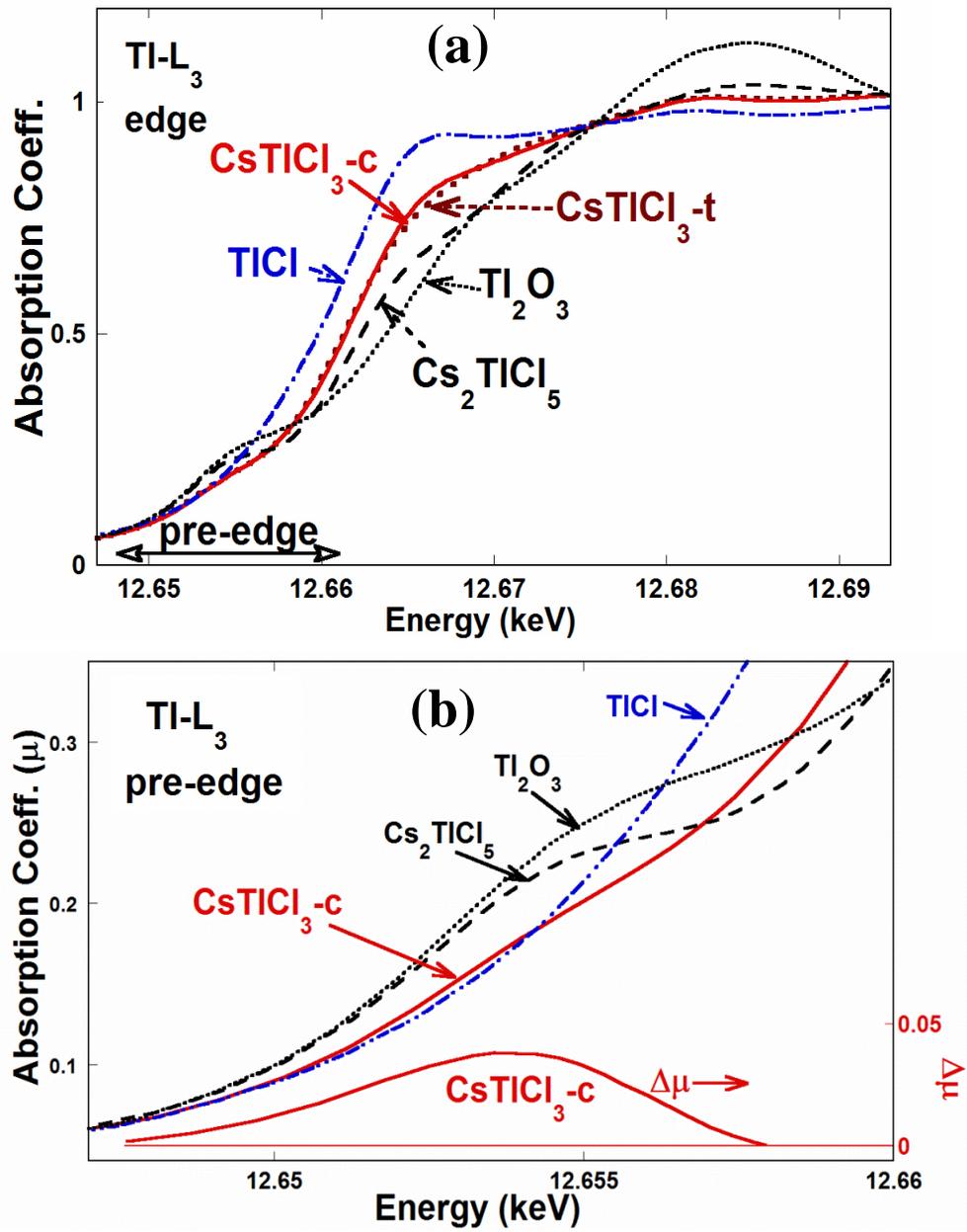

Fig. 6

Retuerto *et al.*

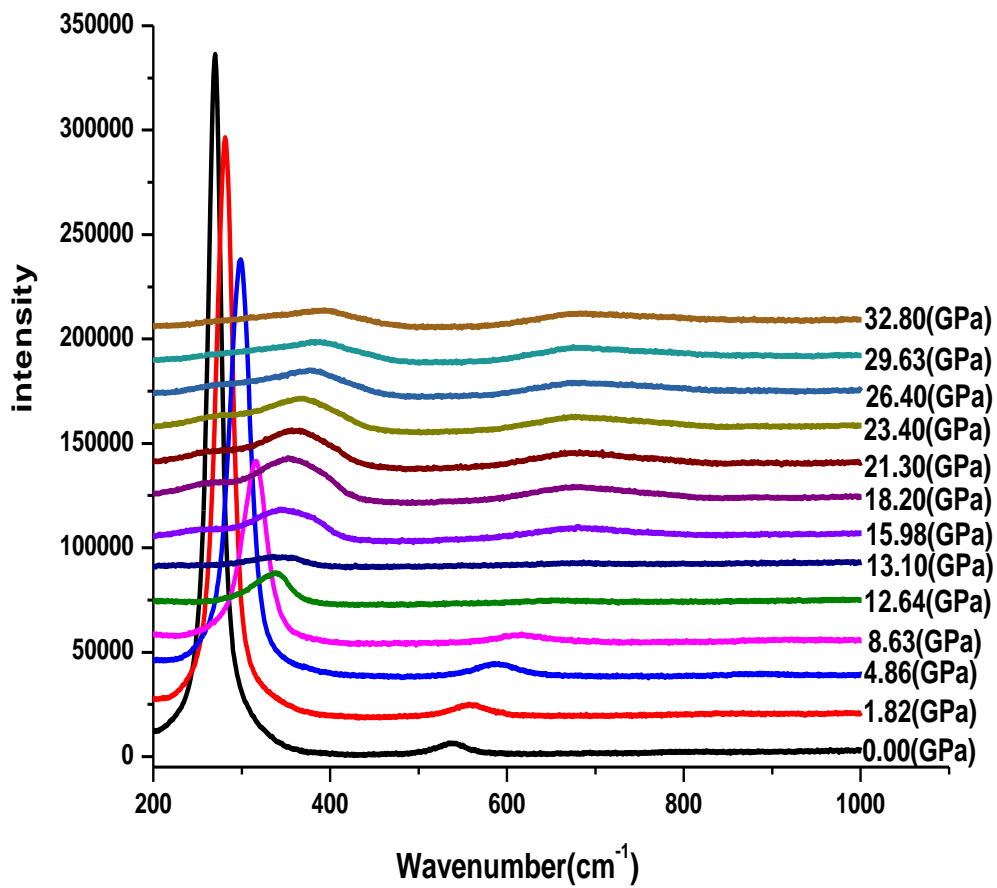

Fig. 7

Retuerto *et al.*

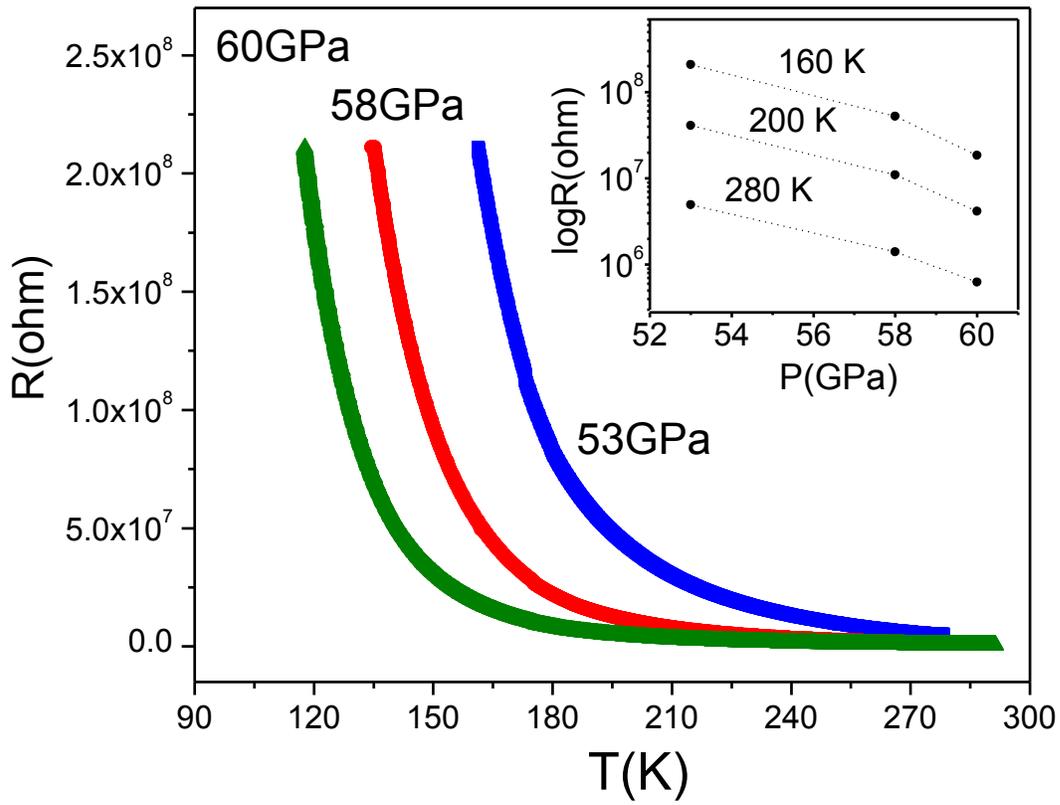

Fig. 8

Retuerto *et al.*

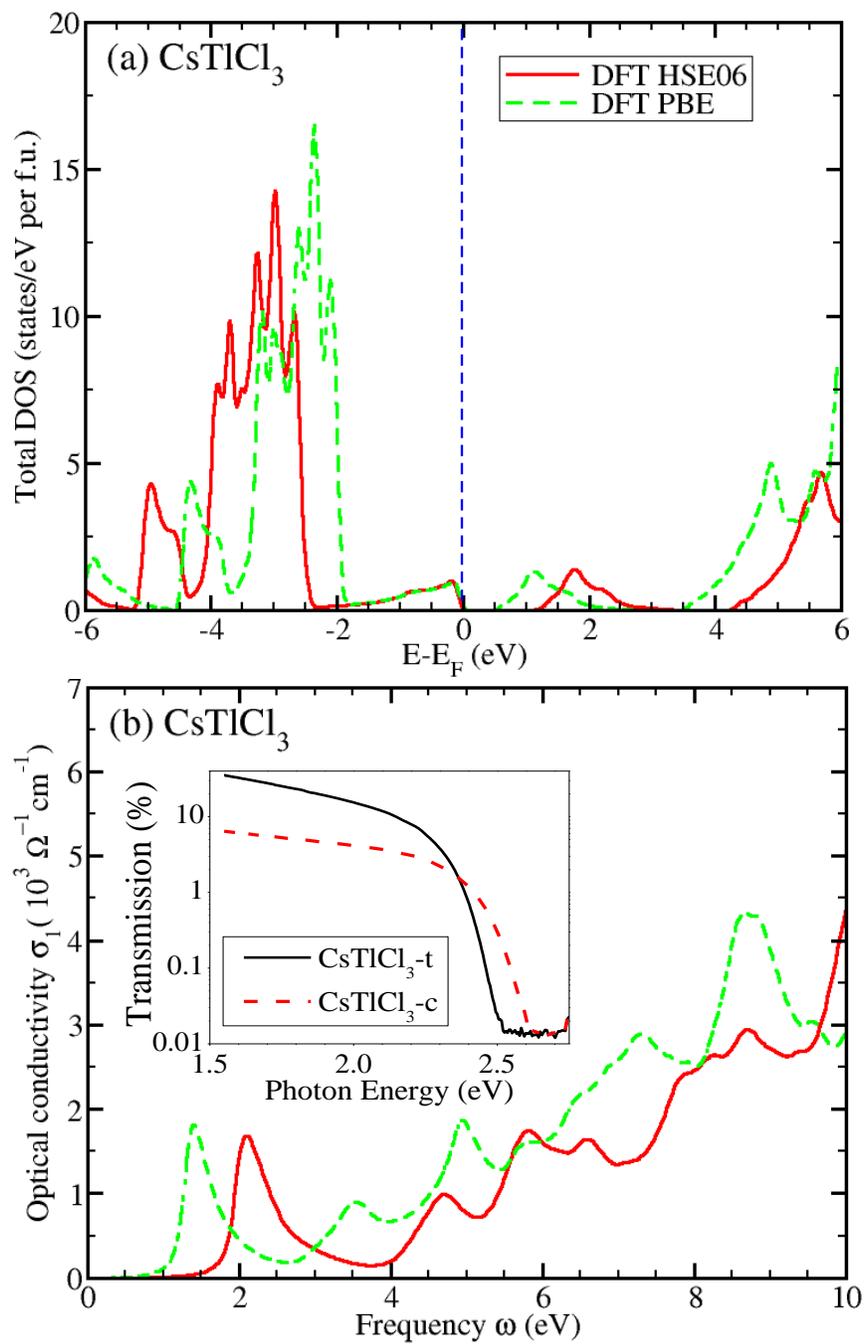

Fig. 9

Retuerto *et al.*

TOC Graphic

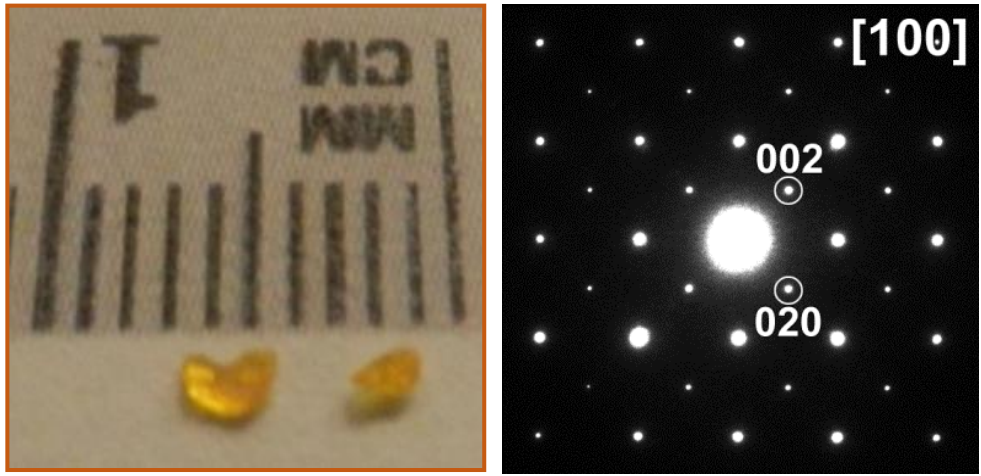
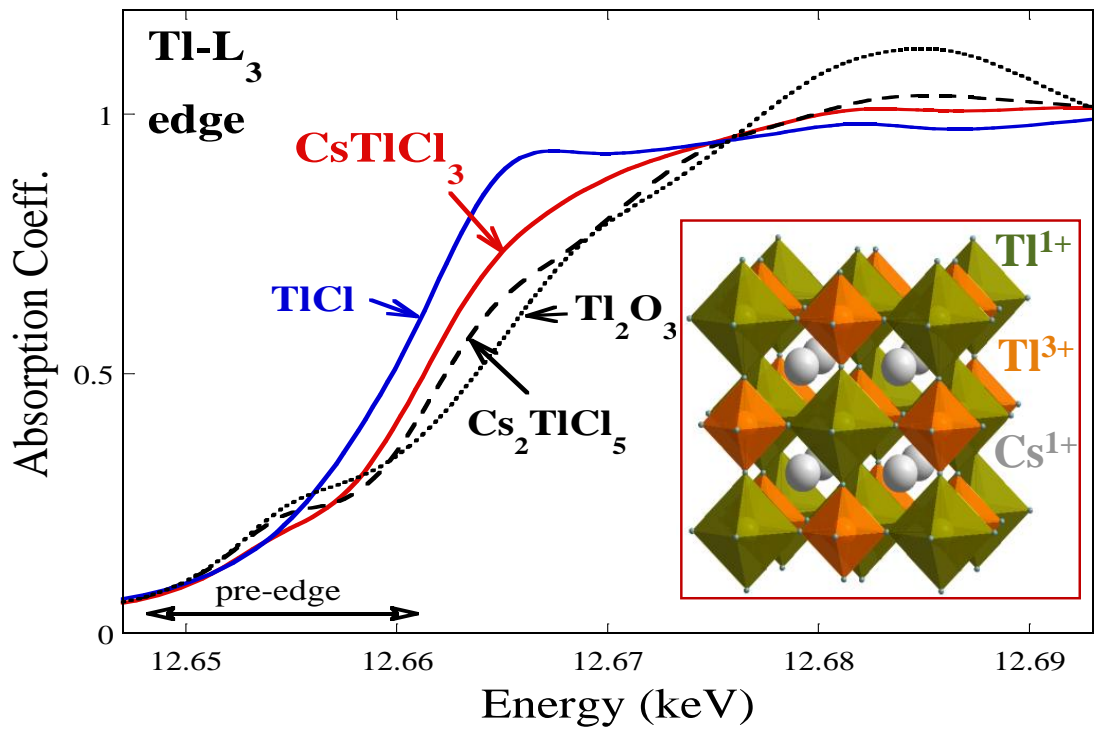